\documentclass{llncs}

\input epsf
\usepackage{graphicx}
\usepackage{url}
\newcommand{\ignore}[1]{}

\begin{document}

\title{Toward a Dynamic Programming Solution \\
       for the 4-peg Tower of Hanoi Problem\\ 
       with Configurations}

\author{Neng-Fa Zhou and Jonathan Fruhman}

\institute{Department of Computer and Information Science, Brooklyn College \\
The City University of New York \\
New York, NY 11210-2889, USA}

\date{}

\maketitle

\begin{abstract}
The Frame-Stewart algorithm for the 4-peg variant of the Tower of Hanoi, introduced in 1941, partitions disks into intermediate towers before moving the remaining disks to their destination.  Algorithms that partition the disks have not been proven to be optimal, although they have been verified for up to 30 disks.  This paper presents a dynamic programming approach to this algorithm, using tabling in B-Prolog.  This study uses a variation of the problem, involving configurations of disks, in order to contrast the tabling approach with the approaches utilized by other solvers.  A comparison of different partitioning locations for the Frame-Stewart algorithm indicates that, although certain partitions are optimal for the classic problem, they need to be modified for certain configurations, and that random configurations might require an entirely new algorithm.
\end{abstract}

\section{Introduction}
The classic 4-peg Tower of Hanoi problem is posed as follows:

Given:  Four pegs, and \emph{n} disks of differing sizes stacked in increasing size order on one of the pegs, with the smallest disk on top.

Goal: Stack the \emph{n} disks on a different one of the four pegs, using the following rules:

\begin{enumerate}
\item
Only one disk can be moved at a time.
\item
Only the top disk on any peg can be moved.	
\item
Larger disks cannot be stacked above smaller disks \cite{ST41}.
\end{enumerate}

A number of authors have created algorithms to try to solve both this puzzle and the related puzzle involving an arbitrary number of \emph{k} pegs.  Frame and Stewart provided the first documented algorithms for the \emph{k}-peg problem \cite{FR41} \cite {ST41}.  Their algorithms involved partitioning the disks into intermediate sub-towers.  Since the time that Frame and Stewart published their results, others have analyzed algorithms in order to either attempt to solve the 4-peg and \emph{k}-peg Tower of Hanoi puzzles using a minimal number of moves, or to try to prove the optimality of the number of moves generated by Frame's and Stewart's algorithms.  Section \ref{sec:fs} discusses the Frame-Stewart algorithm.  

The goal of this paper is to make steps toward an efficient dynamic programming solution for the 4-peg tower of Hanoi puzzle.  The solution is presented using the B-Prolog programming language \cite{Zhou12}, and uses tabling, a technique similar to pattern databases \cite{KO07}, in order to decrease the number of necessary computations.  The program will be \emph{non-deterministic}, in order to extend the Frame-Stewart algorithm to variants of the 4-peg problem that involve disk configurations.  In order to reduce the non-determinism, this paper examines different partition locations for the intermediate sub-towers, in order to find the most efficient one to use for the configuration problem.  Although there is no guarantee that the 4-peg Frame-Stewart algorithm is optimal \cite{LU86}, the dynamic programming solution utilizes this algorithm, as it has been proven to generate the optimal solution for up to 30 disks \cite{KO07}.

In Sect. \ref{sec:asp}, the configuration problem is examined, as posed by the 2011 ASP Competition \cite{TR11}.  The focus is team BPSolver's program \cite{ZH112}, which uses Prolog together with tabling \cite{ZH111}.  BPSolver's results are compared with those of the other teams: team Fast Downward, which used PDDL, and teams EZCSP, IDP, Potassco, and Aclasp, which used SAT solvers and grounders\footnote{https://www.mat.unical.it/aspcomp2011/Participants/}.  The comparison will concentrate on Potassco's program, which had the second-best performance for the Hanoi benchmarks.  Sect. \ref{sec:partitions} compares dynamic programming approaches for splitting the problem into sub-problems, finding one that seems optimal.

Section \ref{sec:conf} studies the seemingly optimal solution when applied to random configurations that do not extend from Frame's and Stewart's algorithms.  Then, in Sect. \ref{sec:classic}, the dynamic programming approaches are applied to the classic problem for up to 30 disks, getting a different result than Sect. \ref{sec:partitions}.  Finally, Sect. \ref{sec:alter} describes alternative approaches to solving the 4-peg and \emph{k}-peg problems.

\section{The Frame-Stewart Algorithm}

\label{sec:fs}

When the \emph{k}-peg Hanoi problem was posed in 1941, two authors provided solutions.  One algorithm, provided by J. S. Frame, is iterative.  The other, written by B. M. Stewart, the proposer, is recursive.  Both authors claim that their algorithms generate the minimum number of moves, while Frame admits that other methods might also be able to provide the same minimum number.
  
Following is Stewart's algorithm for the 4-peg problem:

\begin{enumerate}
\item
Move the $Mid$ topmost disks to another, intermediate, peg (which is not the destination peg), using all 4 pegs.
\item
Move the $n - Mid$ remaining disks to the destination peg using the 3 remaining pegs.
\item
Move the $Mid$ disks from their current peg to the destination peg, using all 4 pegs \cite{ST41}.
\end{enumerate}

Frame's algorithm is similar to Stewart's.  According to Frame, for the \emph{k}-peg problem, a series of towers needs to be created.  The smallest tower will consist of the largest disks, and the largest tower will consist of the smallest disks.  Only the largest disk remains on the start peg.  Once the towers are created, the largest disk is moved directly from the start peg to the destination peg.  Then, the tower of the next-largest disks can be moved to the destination peg using three pegs.  Each intermediate tower is then moved to the destination peg using one more empty peg than the previous tower \cite{FR41}.

If the goal is to minimize the number of disk movements, then this algorithm presents two problems.  One problem is to find the optimal number of disks, $Mid$, to place on the intermediate peg.  This paper examines the partitioning issue in Sect. \ref{sec:partitions}, using Stewart's recursive algorithm.  The other issue is that Frame and Stewart do not provide evidence that an optimal aglorithm involves the creation of intermediate sub-towers \cite{LU86}.  However, as mentioned above, since Korf has verified the optimality of this algorithm for 4 pegs and up to 30 disks \cite{KO07}, the Frame-Stewart algorithm is a good estimate. 

\section{The ASP Competition Problem} 

\label{sec:asp}

\subsection{The 4-Peg Problem with Configurations} 

One of the ASP Competition's problems modifies the 4-peg puzzle to include configurations.  This problem follows the same rules as the classic problem, but differs in regard to how the puzzle is initially arranged, and regarding the goal of the puzzle.  Instead of all the disks beginning and ending on the same peg, this problem provides two \emph{configurations}.  The first configuration is the \emph{start state}, a distribution of the disks over any combination of all four pegs.  The second configuration is the \emph{goal state}, a different distribution of the disks over the four pegs.  These configurations were created based on the moves that the Frame-Stewart algorithm would perform.  The problem is to generate the disk moves needed to get the disks from the first configuration to the second.  In addition, there is a bound on the number of moves that could be generated \cite{TR11}.

The BPSolver team used B-Prolog to generate solutions for the given configurations.  The following is the relevant code:
\begin{verbatim}
  :-table plan4(+,+,+,-,min).
  plan4(N,_CState,_GState,Plan,Len):-N=:=0,!,Plan=[],Len=0.
  plan4(N,CState,GState,Plan,Len):-
     remove_largest_disk_if_in_place(
             N,CState,GState,CState1,GState1),!,
     N1 is N-1,
     plan4(N1,CState1,GState1,Plan,Len).
  plan4(N,CState,GState,Plan,Len):-
     % split disks into two groups
     partition_disks(N,CState,GState,ItState,Mid,Peg),
     % sub-problem1
     remove_larger_disks(CState,Mid,CState1),
     plan4(Mid,CState1,ItState,Plan1,Len1),
     % sub-problem2
     remove_smaller_or_equal_disks(CState,Mid,CState2),
     remove_smaller_or_equal_disks(GState,Mid,GState2),
     N1 is N-Mid,
     plan3(N1,CState2,GState2,Peg,Plan2,Len2),
     % sub-problem3
     remove_larger_disks(GState,Mid,GState1),
     plan4(Mid,ItState,GState1,Plan3,Len3),
     %
     append(Plan1,Plan2,Plan3,Plan),
     Len is Len1+Len2+Len3.
\end{verbatim}

Instead of focusing on the given upper-bound of moves, this program attempts to find the \emph{smallest} number of moves required to get from the start configuration to the goal configuration.  Each configuration is represented by a single state, consisting of four lists.  Each list represents a single peg, and stores the disks that are currently located on the peg.  This is an improvement on the BPSolver team's prior attempts to represent configurations using Boolean expressions.  Prolog lends itself to list operations, and the number of disks on a peg is simply the length of the list.

The predicate {\tt plan4} is defined with three clauses. The first clause is the termination condition.  When the number of disks to be removed is zero, the problem has been solved. The second clause determines whether the current largest disk is currently in place.  If so, it simplifies the problem by logically removing the disk. 

The third clause uses a modified form of Stewart's approach to solve the configuration problem.   The disks are split into two sub-groups, one of which is placed in an intermediate tower.  The problem is separated into three sub-problems: 

\begin{enumerate}
\item
The smallest {\tt Mid} disks are placed on an intermediate peg, {\tt Peg}, by calling {\tt plan4} recursively.
\item
The larger disks are moved from their current pegs to their destination positions by using the deterministic 3-peg algorithm on all of the pegs except for {\tt Peg}\footnote{The 3-peg Tower of Hanoi is deterministic, as it has been proven to have a minimum solution of $2^n - 1$ moves for \emph{n} disks \cite{CH91}.}.
\item
The small disks in the sub-tower are moved from the intermediate peg to their destination positions by calling the {\tt plan4} predicate recursively \cite{ZH111}.
\end{enumerate}

Another important line of code is {\tt :-table plan4(+,+,+,-,min)}.  This uses \emph{tabling} to store information about each state in memory.  The purpose of tabling is to store answers to sub-goals, and to utilize the answers for future variant sub-goals.  This is a useful tool for dynamic programming, which reuses solutions to overlapping sub-problems.  B-Prolog uses linear tabling, lets variant sub-goals share answers, and uses the local, or lazy, strategy to return answers \cite{Zhou08tab}. The most recent version of B-Prolog replies on hash-consing to let tabled subgoals and answers share ground structed terms \cite{ZhouCh12}.

There are two benefits to using tabling.  The first benefit is that tabling prevents infinite loops.  Once a state is visited, it should not be revisited.  If states are visited more than once, the program could be stuck cycling between multiple states, possibly by just repeatedly moving a single disk between two pegs.  By storing states in memory, the program can check the table to see if a state has already been encountered.  The other benefit of using tabling is that it reduces the number of calculations.  Once the program knows how to move \emph{p} disks between two pegs, it can check the table to determine how to move any other set of \emph{p} disks between any two pegs. Sub-problems are represented in such a way that the same problem has the same representation and can share answers through tabling. It does not matter what the sizes of the \emph{p} disks are, nor does it matter which pegs are being used as the current start and destination pegs, assuming the intermediate and destination pegs are logically empty, meaning that they do not contain disks smaller than the largest one being moved.  This decreases the number of operations used during the recursive calls, and is helpful when backtracking to test a different solution.

The line {\tt :-table plan4(+,+,+,-,min)} goes together with the line defining the predicate {\tt plan4(N,CState,GState,Plan,Len)}.  A plus-sign ({\tt +}) indicates that the corresponding arguments ({\tt N} - the number of disks, {\tt CState} - the current state, and {\tt GState} - the goal state) are input.  A minus-sign ({\tt -}) indicates that the corresponding arguments ({\tt Plan} - the sequence of disk moves) are output.  The last part, {\tt min} indicates that {\tt Len}, the length of the plan of disk moves, should be minimized.  By minimizing the plan length, the number of moves will clearly be within the bounds provided in the given problems; otherwise, the problems would be unsolvable.  

Since the predicate {\tt plan4} is non-deterministic, it presents a few interesting issues.  Like the classic algorithm, this program must determine optimal sizes of the disk sub-towers.  A new problem that arises is that the program must determine which peg to use to store each sub-tower.  In the classic 4-peg problem, there is one peg that never has to be used to store a sub-tower \cite{LU89} \cite{VA92}.  However, the same is not true for the configuration problem.  The start and destination pegs can change for each disk, meaning that any peg might need to be used to store a sub-tower.  For further discussion of these issues, see Sect. \ref{sec:partitions}.

Tabling is also used for {\tt plan3}, defined below for the 3-peg problem.

\begin{verbatim}
  :-table plan3(+,+,+,+,-,min).  
  plan3(0,_CState,_GState,_UnusedPeg,Plan,Len):-!,Plan=[],Len=0.
  plan3(N,CState,GState,UnusedPeg,Plan,Len):-
     remove_largest_disk_if_in_place(N,CState,
                         GState,CState1,GState1),!,
     N1 is N-1,
     plan3(N1,CState1,GState1,UnusedPeg,Plan,Len).
  plan3(1,CState,GState,_UnusedPeg,Plan,Len):-!,
     Plan=[(Peg1,Peg2)],Len=1,
     btm_disk_on_peg(1,CState,Peg1),
     btm_disk_on_peg(1,GState,Peg2).
  plan3(N,CState,GState,UnusedPeg,Plan,Len):-
     btm_disk_on_peg(N,CState,Peg1),
     btm_disk_on_peg(N,GState,Peg2),
     other_two_pegs(Peg1,Peg2,Peg3,Peg4),
     (UnusedPeg==Peg3->TmpPeg=Peg4;TmpPeg=Peg3),
     N1 is N-1,
     remove_bottom_disk(CState,Peg1,CState1),
     ItState=s(_,_,_,_),
     Tower @= [I : I in N1..(-1)..1],
     foreach(I in 1..4, 
             (I==TmpPeg->arg(I,ItState,Tower);
              arg(I,ItState,[])
             )
     ),
     plan3(N1,CState1,ItState,UnusedPeg,Plan1,Len1),
     remove_bottom_disk(GState,Peg2,GState1),
     plan3(N1,ItState,GState1,UnusedPeg,Plan2,Len2),
     append(Plan1,[(Peg1,Peg2)|Plan2],Plan),
     Len is Len1+Len2+1.
\end{verbatim}

Like {\tt plan4}, the predicate {\tt plan3} has a clause for the termination condition and a clause for reducing the problem when the latest disk is in place. When the number of disks to be moved is one, it takes one step to solve it.  Otherwise, the problem is divided into three sub-tasks: the first sub-task is to move N-1 disks except for the largest one from the current peg to a temporary peg; the second sub-task is to move the largest disk to the destination peg; the third task is to move the N-1 disks from the temporary peg to the destination peg. Unlike {\tt plan4}, {\tt plan3} is deterministic.

\subsection{Competition Results}

The ASP Competition had six participants, including BPSolver\footnote{The following implementation details, and the participants' programs, can be obtained at the team description pages, located at https://www.mat.unical.it/aspcomp2011/Participants/.  Further details can be found at the solvers' websites, as listed in the teams' descriptions online.}.  FastDownward used Planning Domain Definition Language to represent the problems, and utilized A* search with the selective-max and landmark-cut heuristics,  using the input to bound the maximum solution length, in order to solve the Hanoi problem.  The IDP team described the problems by using First Order Logic with Inductive Definitions, translated the problems into Extended Conjunctive Normal Form by using the Gidl grounder, and solved the problems with the MINISAT(ID) solver.  The remaining three teams solved the Hanoi problem by using the grounder Gringo, which translates input programs into equivalent, variable-free programs, and the solver Clasp, which focuses on answer set programming together with nogood learning, checking for violated constraints.  While EZCSP ran Clasp on its default settings, Aclasp utilized a modified restart strategy, which, as described by the team, ``depends on the average decision-level on which conflicts occurred.''  In order to solve the Hanoi problem, the final team, Potassco, specified that Clasp should use the Variable State Independent Decaying Sum decision heuristic, and limited the amount of preprocessing and the initial database size.

The ASP Competition programs were graded based on two criteria: the correctness of the solution and the amount of time that it took for each program to find the solution.  The competition organizers had sixty  instances\footnote{\url{http://www.mat.unical.it/aspcomp2011/files/HanoiTower/hanoi_tower-full_package.zip}} that could have been used to test the Hanoi solvers.  Only fifteen of those instances were actually used to grade the solvers.  Based on those fifteen instances, the BPSolver team scored highest, followed by Potassco, AClasp, and IDP.  EZCSP and Fast Downward tied for the lowest score.

BPSolver's program was the only one that actively used the Frame-Stewart algorithm to limit the size of the search space.  Every other program defined legal moves for the tower of Hanoi problem, and used the definition as input to their solvers.  Therefore, the other teams' scores were a result of the solvers they used and the heuristics that the solvers employed to trim the search space.  The active use of the Frame-Stewart algorithm contributed to the speed of BPSolver's program.  Although some of the other teams used heuristics, and added rules in order to improve the search, such as EZCSP specifying that no disk should be moved twice in a row, they still had a larger search space.  As will be shown in Sect. \ref{sec:partitions}, BPSolver used a mathematical function to partition the disks, which limited the possible moves at any given time.  In addition, tabling reduced the number of repeated calculations, causing BPSolver's program to run faster than all the other programs.

In the competition, BPSolver's Hanoi program scored 94, while Potassco's scored 81.  However, an analysis of their performances on all 60 instances, summarized in Table \ref{tab:comp}, shows that the margin would not be as large.  BPSolver still scores higher than Potassco, but that is primarily due to the speed of B-Prolog, as opposed to the correctness of the program.

\begin{table}
\caption{\label{tab:comp}Comparing B-Prolog's original program with Potassco's program}
\resizebox{\textwidth}{!}{%
\begin{tabular}{|c|c|c|c|c|c|c|}
\hline
\underline{\textbf{Sys}} & \underline{\textbf{AVG Seconds}} & \underline{\textbf{MAX Seconds}} & \underline{\textbf{Num Solved}} & \underline{\textbf{SSolve}} & \underline{\textbf{STime}} & \underline{\textbf{Score}} \\
\hline
BPSolver & 0.022883333 & 0.047 & 54 & 45 & 45 & 90 \\
\hline
Potassco & 10.61766667 & 102.867 & 60 & 50 & 37 & 87 \\
\hline
\end{tabular}}
\end{table}

BPSolver's program printed a result of ``UNKNOWN'' for 6 of the instances, while Potassco's was able to solve all instances.  Four of those six instances are a result of an error\footnote{When a sub-tower is built, it should be built on a logically empty peg.  The original program did not test if the peg was empty, causing the error.}, in which the program printed ``UNKNOWN'' after finding a solution.  Only one of these instances was run during the competition.  The remaining two unsolved instances, which are illustrated in Fig. \ref{fig:unsolved_conf}, and which were not run during the competition, are indicative of a larger problem.  Team BPSolver mistakenly believed that there was a known optimal guide for partitioning the disks when creating intermediate sub-towers \cite{ZH111}.  The B-Prolog program's partitioning guide was not always able to find an optimal solution.

\begin{figure}
\centering
\includegraphics[scale=0.45]{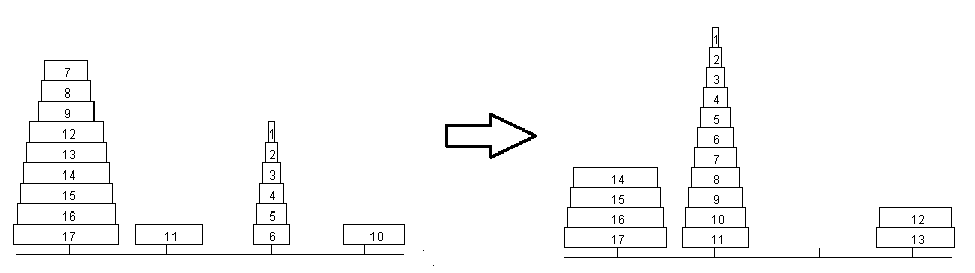}
\includegraphics[scale=0.48]{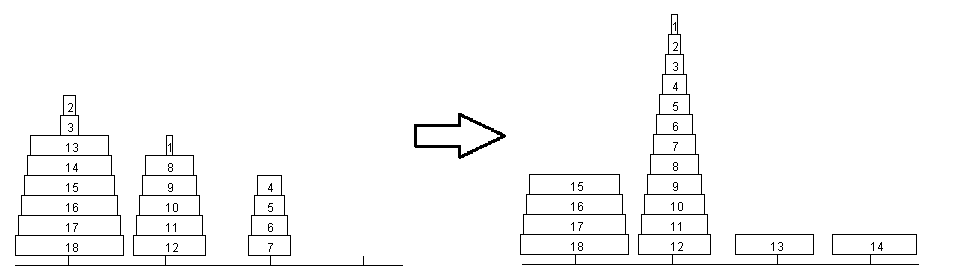}
\caption{\label{fig:unsolved_conf}Instances 17 and 22, the unsolved instances}
\end{figure}
\ignore{
\begin{center}
\begin{figure}
\epsfxsize=6cm 
\centering{\epsfbox{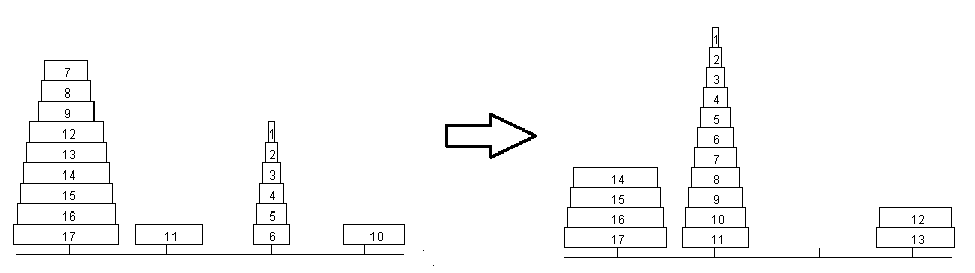}}
\centering{\epsfbox{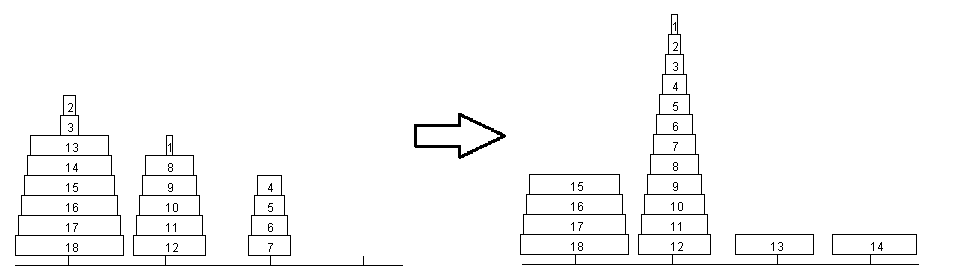}}
\caption{\label{fig:unsolved_conf}Instances 17 and 22, the unsolved instances}
\end{figure}
\end{center}
}

\subsection{Partitions}
\label{sec:partitions}

There have been a number of different estimates for determining the optimal size of the intermediate sub-tower when given \emph{n} disks.

For the ASP competition, BPSolver used the following code to determine the optimal partition location:
\begin{verbatim}
  partition_disks(N,CState,GState,ItState,Mid,Peg) :-
     btm_disk_on_peg(N,CState,Peg1),
     btm_disk_on_peg(N,GState,Peg2),
     other_two_pegs(Peg1,Peg2,Peg3,Peg4),
     (Peg=Peg3;Peg=Peg4),
     arg(Peg,CState,CDisks),
     arg(Peg,GState,GDisks),
     PN is N-integer(sqrt(2*N)+0.5),
     Low is max(PN-2,1),
     Up is min(PN+2,N-1),
     between(Low,Up,Mid),
     (CDisks=[CBDisk|_],CBDisk=<Mid; CDisks=[]),
     (GDisks=[GBDisk|_],GBDisk=<Mid; GDisks=[]),
     Tower @= [I : I in Mid..(-1)..1],
     ItState=s(_,_,_,_),
     foreach(I in 1..4,
             (I==Peg->arg(I,ItState,Tower);
              arg(I,ItState,[])
             )
     ).
\end{verbatim}

The arguments are similar to the ones in the {\tt plan4} predicate, except {\tt ItState} defines the intermediate state with the sub-tower, and {\tt Peg} is the peg on which the sub-tower will be placed.

As mentioned above, there is an additional issue about where to place the sub-tower, based on the configurations.  The program solves it in one of two ways.  If the current configuration has a pre-existing sub-tower, a new one does not need to be created.  Otherwise, the program non-deterministically tests both of the pegs that are currently serving as intermediate pegs for the optimal sub-tower location.

Given \emph{n} disks, the program uses Rand's estimate of $n - \lfloor \sqrt{2n} + 0.5 \rfloor$ for the sub-tower size.  As discovered after the ASP Competition, this does not always generate the optimal solution.  Therefore, the program was tested using different partition estimates.  Table \ref{tab:part} shows the results\footnote{All of the tests described in this paper used B-Prolog version 7.5, which is the version used for the competition.  Later versions of B-Prolog perform tabling in a different manner, decreasing the amounts of used table memory in some cases, and increasing the amounts used in other cases.}.

\begin{table}
\caption{\label{tab:part}Partitions}
\resizebox{\textwidth}{!}{%
\begin{tabular}{|c|c|c|c|c|c|c|}
\hline
\underline{\textbf{Partition}} & \underline{\textbf{Source}} & \underline{\textbf{AVG Seconds}} & \underline{\textbf{AVG Table Used (Bytes)}} & \underline{\textbf{MAX Seconds}} & \underline{\textbf{MAX Table Used (Bytes)}} & \underline{\textbf{UNKNOWN}} \\
\hline
$N - \lfloor \sqrt{2 * N} + 0.5 \rfloor$ & \cite{RA09} & 0.0234 & 121135.4 & 0.047 & 207724 & 2 \\
\hline
$N - \lceil \sqrt{2 * N + 0.25} - 0.5 \rceil$ & \cite{VA92} & 0.0209 & 121135.4 & 0.032 & 207724 & 2 \\
\hline
$N - \lfloor \frac{\sqrt{8 * N + 1} - 1}{2} \rfloor$ & \cite{CH91} \cite{RO86} \cite{WA07} & 0.023 & 141369.4 & 0.032 & 206728 & 0 \\
\hline
$N - K$, where $K$ is the smallest integer & \cite{ZH08} & 0.021566667 & 122005.1333 & 0.032 & 208868 & 2 \\
such that $T_K \ge N$  (with tabling of results). & & & & & & \\
\hline
$N - K$, where $K$ is the largest integer & \cite{CH91} & 0.02285 & 142401.9333 & 0.032 & 207960 & 0 \\
such that $T_K \le N$  (with tabling of results). & & & & & &\\
\hline
$N - K$, if  $N = T_K$ for some $K$ & \cite{ST94} & 0.025833333 & 162988.3333 & 0.047 & 237360 & 0 \\
Otherwise, N - K, or N - (K + 1), & & & & & & \\
where K is the largest integer & & & & & &\\
such that $T_K < N$  (with tabling of results). & & & & & &\\
\hline
Program Decides Without a Guide & & 0.039466667 & 4457510.067 & 0.124 & 16884172 & 0 \\
\hline
\end{tabular}}
\end{table}

For Table \ref{tab:part}, two types of estimates were used.  One type is a formula stated in terms of the number of disks.  The other type relates the 4-peg Hanoi problem to the \emph{triangular numbers}, which are of the form $T_k = \frac{k * (k + 1)}{2}$ \cite{RO86}, by finding the triangular number that is closest to the number of disks.  The estimates of the first type explicitly state the relation between the number of disks and the triangular numbers found by the second type \cite{CH91}.  In the final row of Table \ref{tab:part}, the program was not given a guide for where to partition the disks.  Instead, it tested every number between 1 and $n - 1$, where \emph{n} is the number of disks.

There are a number of other guides for partitioning the disks \cite{KL02} \cite{LU89} \cite{MA96} \cite{ST94}, but these were not tested.  Some, such as \cite{LU89}, are too closely related to specific algorithms.  Others, like \cite{KL02}, are too mathematically complex to test in terms of this program.  Some of the remaining, like \cite{MA96}, are for specific cases.  Some alternative partitions are clearly incorrect, as shown by Stockmeyer \cite{ST94}.  Instead of including Frame's and Stewart's estimates, this study incorpates Stockmeyer's estimate, which is based on those of Frame and Stewart \cite{ST94}.

The B-Prolog program allows for an error in the number, $PN$, returned by the estimates.  It checks all values between $PN - 2$ and $PN + 2$ for an optimal partition number.  If this error bound is removed, and the program uses the exact number that is found, or if the error bound is decreased to $PN - 1$ and $PN + 1$, then none of the functions in Table \ref{tab:part} solve all of the instances.  It should be noted that three of the estimates in Table \ref{tab:part} fail for two of the instances, 17 and 22, even when the error is allowed.

The different approximations were tested for time and the amount of table memory that they used.  On average, the explicit functions almost always used less memory than the partitions that calculated each triangular number until a certain condition was met.  The maximum memory used was always smaller for the explicit functions than for the repeated calculations.  One reason for this is that tabling was incorporated into the calculations of the triangular numbers in order to reduce the number of required calculations.  As shown in Table \ref{tab:part}, if the program tests every possible partition number, then it is guaranteed to find the optimal solution.  However, the time and memory costs are much higher than they are if the program is given a partitioning guide.

Table \ref{tab:part} indicates that the best guide to use for the competition is the one originally provided by Rohl and Gedeon (as shown in row 3).  It should be noted that, if floor and ceiling operations are removed, Liefvoort's solution (row 2) is mathematically equivalent to theirs.  Since Rohl and Gedeon use the floor operation, while Liefvoort uses the ceiling operation, their solutions differ.  

After the original program's error was removed, and the partitioning estimate was changed to that of Rohl and Gedeon, BPSolver's modified program was run on all 60 instances.  Table \ref{tab:new_comp} shows the results of these tests, as compared to Potassco's results.  The results show that Rohl's and Gedeon's numbers are better than Rand's, because BPSolver now scores 100, instead of 94.

\begin{table}
\caption{\label{tab:new_comp}Comparing B-Prolog's modified program with Potassco's program}
\resizebox{\textwidth}{!}{%
\begin{tabular}{|c|c|c|c|c|c|c|}
\hline
\underline{\textbf{Sys}} & \underline{\textbf{AVG Seconds}} & \underline{\textbf{MAX Seconds}} & \underline{\textbf{Num Solved}} & \underline{\textbf{SSolve}} & \underline{\textbf{STime}} & \underline{\textbf{Score}} \\
\hline
BPSolver & 0.023 & 0.032 & 60 & 50 & 50 & 100 \\
\hline
Potassco & 10.61766667 & 102.867 & 60 & 50 & 37 & 87 \\
\hline
\end{tabular}}
\end{table}

\section{Random Configurations and the Classic Problem}

Each of the sixty instances generated for the 2011 ASP Competition consisted of configurations that would be created during the execution of the Frame-Stewart algorithm on the classic 4-peg problem.  The configurations included pre-created sub-towers that the Frame-Stewart algorithm would have generated.  This section will show that, although a certain partition number is shown to work on the ASP configurations, it is not guaranteed to solve every possible random configuration.  

BPSolver's modified program was tested in two additional ways.  It was run on randomly generated configurations, some of which might not be created during the execution of the Frame-Stewart algorithm, and it was run on the classic 4-peg problem.

\subsection{Random Configurations}

\label{sec:conf}

The first test was running the BPSolver program on randomly generated configurations.  They were generated to match the ASP Competition's input files, which consisted of five sets of predicates \cite{TR11}.
\ignore{
\begin{enumerate}
\item
{\tt step(N).} \emph{N} is an integer representing the maximum number of steps that were allowed.  
\item
{\tt time(0). time(1). ... time(N).} There were \emph{N + 1} time predicates, where each \emph{N} represented a single step, and the last one represented the final configuration.  
\item
{\tt disk(1). ... disk(D + 4).} \emph{D} is the number of disks.  The first four disk predicates represented the four pegs, and the remaining \emph{D} represented the disks, where larger numbers of \emph{D} indicate the smaller disks\footnote{For this paper, images will differ from the program.  The images will use smaller numbers to represent the smaller disks}.  
\item
A series of predicates {\tt on0(A, B)}, indicating that in the starting configuration, disk \emph{A} is directly on top of disk (or peg) \emph{B}. 
\item
A series of predicates {\tt ongoal(A, B)}, indicating that in the final configuration, disk \emph{A} will be directly on top of disk (or peg) \emph{B}.
\end{enumerate}
}
The BPSolver program was tested on twenty random configurations.  Since the configurations were random, some appeared to contain pre-existing intermediate sub-towers, while others did not.  Two versions of BPSolver's program were tested.  One version created the sub-towers using the formula originally posed by Rohl and Gedeon \cite{RO86} with an error bound of two, and the other did not use any guide for where to partition the disks.

These tests had varied results.  Only twelve of the twenty instances were solved by both programs.  These instances had between 18 and 20 disks, while their solutions required between 120 and 257 steps.  Another five instances were only solved by the program that did not use a partitioning guide.  These instances had between 21 and 23 disks, and required between 387 and 408 steps.  One instance, with 25 disks, was solved in 533 steps by the program that used a formula, while the program that did not use a guide ran out of memory.  The remaining two instances, which had 24 and 25 disks, were solved by neither program.

The results appear to indicate that the 4-peg problem with random configurations requires a different approach than that of the Frame-Stewart algorithm.  Due to the varying sizes of the disks on any peg in the starting configuration, it might not be easy to derive a mathematical formula for the creation of an intermediate sub-tower.  If a guide is not used, it can be too computationally complex to test every possible set of moves.  Disks may need to be repeatedly moved between the four pegs before a condition arises in which it is possible to create an intermediate sub-tower.  Configurations such as those found in Fig. \ref{fig:rand_conf} might not be encountered during the regular execution of the Frame-Stewart algorithm.  Therefore, unless modifications are introduced to the algorithm, the Frame-Stewart algorithm might not be the best to solve random configurations.

\begin{figure}
\centering
\includegraphics[scale=0.35]{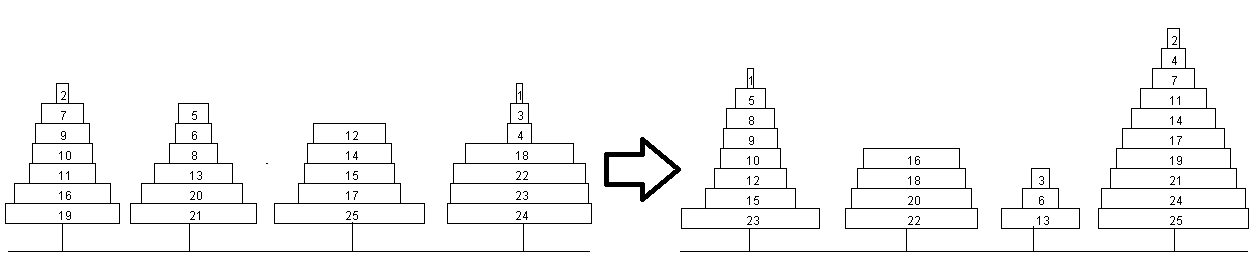}
\caption{\label{fig:rand_conf}Start and end configurations of an unsolved random instance}
\end{figure}
\ignore{
\begin{center}
\begin{figure}
\epsfxsize=6cm 
\centering{\epsfbox{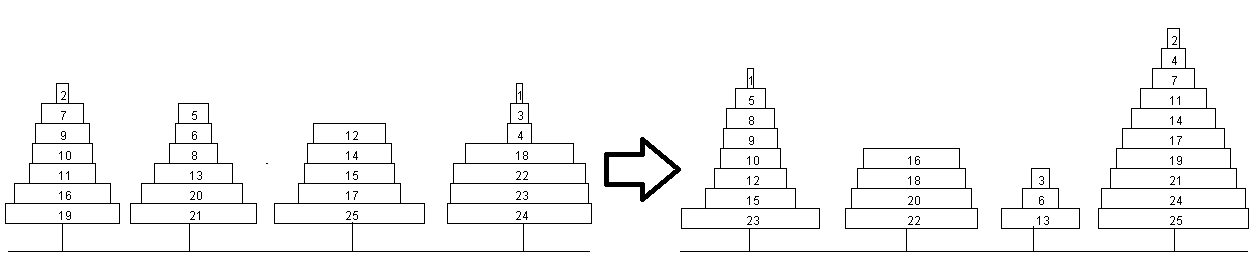}}
\caption{\label{fig:rand_conf}Start and end configurations of an unsolved random instance}
\end{figure}
\end{center}
}

\subsection{The Classic 4-peg Problem}

\label{sec:classic}

The classic 4-peg problem was used to test the same two programs that were tested on the random configurations.  For this problem, the programs were tested using an input of 30 disks.  The program that used a formula was able to solve the problem using the optimum possible number of moves, 1025 \cite{KO07}.  The one that had no guide ran out of memory.

Since the program that used Rohl's and Gedeon's guide was able to solve the 30-disk problem, it was tested on the classic problem using every number of disks between one and thirty.  When these tests were run, the partitioning error bounds of +2 and -2 were removed.  For each of the thirty instances, the program was able to solve the problem using the optimum possible number of moves, as proven by \cite{KO07} and \cite{KO05}.

After these tests were performed, the classic 4-peg problem was used to test the same six partition estimates that were examined using the ASP problem in Table \ref{tab:part}.  Each estimate was tested on every number of disks between one and thirty.  As opposed to the prior tests, these tests did not allow any error bound when calculating the partition number.  Although three of the estimates failed to solve every single ASP Competition configuration, all six estimates generated optimal solutions for every instance of the classic problem.  Korf has verified the minimum possible number of moves for any algorithm using inputs of up to 30 disks.  By using the Frame-Stewart algorithm with any of the six partition estimates on the regular 4-peg Hanoi problem, a solution is obtained that is the optimal solution of any possible algorithm.  Tabling and testing every partition number can be too computationally complex, but Rohl's and Gedeon's formula provides a good guide.

\section{Alternative Approaches}

\label{sec:alter}

Many authors have tried to solve the 4-peg and \emph{k}-peg problems using their own algorithms.  Although most of the approaches are not used in this paper, it is important to note their contributions.  Following is a subset of the algorithms:

\subsubsection{Rohl's and Gedeon's Algorithm.}
 
Rohl and Gedeon created recursive algorithms for both the 4-peg and the \emph{k}-peg problems, using a form of Stewart's algorithm.  This paper will focus on the algorithm for four pegs, wherein each recursive call builds a single sub-tower.  If the current ordering of the pegs is (1, 2, 3, 4), the 4-peg algorithm is called to create a sub-tower with ordering (1, 4, 3, 2), the 3-peg algorithm is called on (1, 2, 3) to move the remaining pegs, and the 4-peg algorithm is called on (4, 3, 2, 1) to move the sub-tower to the destination peg \cite{RO86}.  This algorithm provided the partitioning guide utilized by BPSolver's modified program.

\subsubsection{Lu's Algorithm.}

Lu produced an iterative approach for the 4-peg problem.  Lu's algorithm shows a correlation between the disk moves and binary numbers.  It inserts a number of logical fake disks at pre-determined locations in order to map the disk moves to the binary numbers \cite{LU89}.

\subsubsection{Sarkar's Algorithm.}

Sarkar used a recursive algorithm for the \emph{k}-peg problem that is similar to Frame's algorithm.  The algorithm deterministically decides how to distribute the topmost disks over intermediate pegs, using summation functions, before distributing any disks.  It then recursively calls itself to distribute the disks.  Sarkar posed the \emph{serialization conjecture}, stating that an optimal algorithm will distribute disks onto pegs such that disks on each peg have consecutive sizes, and the size of the top disk on a peg is consecutive with the size of the bottom disk on the next peg.  If this conjecture is valid, then Sarkar's algorithm is optimal \cite{SA00}.  Although this distribution method is efficient, this paper does not use it, because it does not easily extend to the configuration problem.

\subsubsection{Wang's, Liu's, Yue's, Shao's, and Lu's Algorithm.}
	
Another iterative algorithm for the 4-peg puzzle involves arranging the disk numbers into an upper-triangular array.  Based on this array, there is a ``cross-correlation'' between the solution of the problem represented by location (\emph{i}, \emph{j}) and the problems represented by locations (\emph{i}, \emph{j+1}), and (\emph{i+1}, \emph{j+1}).  This correlation decreases the number of necessary calculations \cite{WA07}.  This is another efficient algorithm, although there is no clear way to extend it to the configuration problem.

\section{Summary}

For configurations directly based on the Frame-Stewart algorithm, Rohl's and Gedeon's partitioning estimate of $N - \lfloor \frac{\sqrt{8 * N + 1} - 1}{2} \rfloor$ appears to be an optimal guide, even if it requires error bounds.  Although all of the estimates in Table \ref{tab:part} generate optimal solutions for the regular 4-peg puzzle for up to 30 disks without using error bounds, their correctness does not seem to extend to the configuration problem.  Perhaps the nature of the configuration problem changes the optimal partition location.

Testing BPSolver's program on random configurations further clarifies the issue.  It appears that there are some configurations for which creating intermediate sub-towers may not be optimal.  Unlike the problem as studied by Frame and Stewart, it is not clear where to place the intermediate sub-towers.  The non-determinism involved can cause the computation problem to be too computationally complex.  However, BPSolver's program with Rohl's and Gedeon's estimates runs quickly, and uses tabling to reduce the number of calculations.  Therefore, it is a good guide for a dynamic programming solution for modified forms of the 4-peg Tower of Hanoi problem where non-determinism is required.

The B-Prolog program demonstrates the importance of tabling for declarative description of dynamic programming solutions. Like use of pattern databases in state-space search and conflict-driven clause learning and memoization in SAT solvers, tabling is a great technique for avoiding repeated exploration of the same states during search.

\section*{Acknowledgements}
This research was supported in part by NSF (No.1018006)

\end{document}